\documentstyle[prd,aps,floats]{revtex}

\newcommand{\be}{\begin{equation}}
\newcommand{\ee}{\end{equation}}
\newcommand{\bea}{\begin{eqnarray}}
\newcommand{\beas}{\begin{eqnarray*}}
\newcommand{\eea}{\end{eqnarray}}
\newcommand{\eeas}{\end{eqnarray*}}
\newcommand{\ba}{\begin{array}}
\newcommand{\ea}{\end{array}}
\newcommand{\bi}{\begin{itemize}}
\newcommand{\ei}{\end{itemize}}
\newcommand{\ben}{\begin{enumerate}}
\newcommand{\een}{\end{enumerate}}
\def\ls{\mathrel{\lower4pt\vbox{\lineskip=0pt\baselineskip=0pt
           \hbox{$<$}\hbox{$\sim$}}}}
\def\gs{\mathrel{\lower4pt\vbox{\lineskip=0pt\baselineskip=0pt
           \hbox{$>$}\hbox{$\sim$}}}}
\begin{document}
\draft

\input epsf \renewcommand{\topfraction}{0.8} 
\twocolumn[\hsize\textwidth\columnwidth\hsize\csname 
@twocolumnfalse\endcsname

\title{Affleck-Dine baryogensis in large extra dimensions} 
\author{Anupam  Mazumdar$^a$ and A. P\'erez-Lorenzana$^{a,b}$}
\address{
$^a$ The Abdus Salam International Centre for Theoretical Physics, I-34100,
Trieste, Italy \\
$^b$ Departamento de F\'{\i}sica,
Centro de Investigaci\'on y de Estudios Avanzados del I.P.N.\\
Apdo. Post. 14-740, 07000, M\'exico, D.F., M\'exico}
\date{March 2001} 
\maketitle

\begin{abstract}
Baryogenesis in the models where the fundamental scale is as low as 
TeV in the context of large extra dimensions is a challenging problem. 
The requirement for the departure from thermal equilibrium necessarily
ties any low scale baryogenesis with that of a successful 
inflationary model which automatically provides the out of equilibrium 
condition after the end of inflation. However, it is also noticeable 
that in these models the reheat temperature of the Universe is 
strongly constrained from the overproduction of Kaluza-Klein modes,
which enforces a very low reheat temperature. In this paper 
we describe a possible scenario for baryogenesis which has 
a similar characteristics of an Affleck-Dine field. We notice that 
in order to have an adequate baryon to entropy ratio one requires to 
promote this Affleck-Dine field to reside in the bulk. 
\end{abstract}

\vspace*{-0.6truecm}
\vskip2pc]


\section{introduction} 
\vspace*{-0.2truecm}
Baryogenesis is an interesting offshoot of cosmology and particle 
physics, which tries to explain why the ratio of baryon density 
and photon density is given by one part in $10^{10}$ during the 
nucleosynthesis era~\cite{ratio}. The synthesis of light elements
depends crucially on this ratio which tells us that in an absence of 
any observed anti-matter region, the baryon density should be equal
to the cosmological baryon asymmetry. There are many proposals which
can satisfy the three conditions; namely $C$ and $CP$ violation, $B$ or $L$ 
violation, and, out of equilibrium decay, which are the essential ingredients
for baryogenesis \cite{sakh}. Out of the three mentioned
conditions the last one has to come purely from the cosmological evolution
of the Universe. It is quite probable that the early Universe might 
have had strong departure from thermal equilibrium due to a large 
expansion rate of the Universe and presence of heavy decaying particles, 
however, this possibility gradually becomes difficult to acquire at 
scales which are comparable to the electroweak scale. 
As a second alternative, one might expect to attain the 
departure from thermal equilibrium via some phase transitions which
would break global or gauge symmetry, a perfect example is electroweak
phase transition 
where there is  an anomalous $B+L$ violation,
 for a review, see Ref.~\cite{buchmuller}. 
In the former situation, the departure from thermal equilibrium is 
usually tied up with inflation.
Inflation is an attractive paradigm which solves a range of troublesome
problems of the Big Bang cosmology besides acting as the best candidate 
for producing an almost scale invariant density perturbations. After
a period of inflation the Universe undergoes through an era of reheating, 
and this is precisely where one might expect to produce
massive bosons and their out of equilibrium decay which might lead
to the desired baryon to entropy ratio.

On the other hand recent trends in solving the hierarchy problem,
in the context of theories with extra dimension, 
suggest that the strength of the fundamental scale, might be much 
lower than the four dimensional Planck scale. If that scale be
the electroweak scale then the hierarchy between the Planck scale and
the electroweak scale can be inverted by assuming that there exists 
large extra dimensions, which can be as large as mm~\cite{nima}. 
It is also assumed that the SM
particles are trapped in a four dimensional hypersurface (a $3$-brane), 
thus, they are not allowed to propagate in the bulk.
However, it is generically assumed that besides
gravity, SM singlets may propagate in the bulk.
Among them the inflaton can be a candidate, which is 
less favored to be a brane field 
(see for instance Refs. \cite{abdel0,abdel1}). 
However, in these models
the Universe during the radiation dominated epoch reaches its maximum
temperature very close to MeV which we shall discuss in details in the
coming sections. For such a low reheat temperature baryogenesis is a 
challenging task because of two reasons; first of all the late decay of 
particles including the inflaton which is responsible for reheating the 
Universe, and, secondly the operators which might lead to baryon number
violation must be suppressed due to stringent constraints on 
proton life time. This restricts us to a few choices of baryogenesis 
models which may work well in presence of a small 
fundamental scale, such as $\sim {\cal O}(\rm TeV)$ \cite{davidson0}.

Other possibility may appear from the fact that reheat temperature
is not the maximum temperature in the Universe after the end of
inflation. Usually, reheating takes a while and it is possible to reach
a temperature during the process of reheating which can be quite large,
however, this rise in temperature crucially depends on the scale when 
inflation comes to an end \cite{giudice}. 
 If this be the case, then, it is quite possible 
that the rate of sphaleron transitions are active, even though 
the reheat temperature is much smaller than $100$ GeV \cite{davidson1}. 
In this paper we describe a completely different possibility. This 
mechanism does not
depend on the predictability of high rise in temperature during the
reheating era. Our scheme is analogous to the Affleck-Dine (AD) mechanism 
of baryogenesis. We organize our paper with a brief discussion on 
reheat temperature and the bounds upon the reheat temperature, then
we describe the possibility of leptogenesis which can be reprocessed
into baryon number $B$ by anomalous $B+L$ violating sphaleron interactions,
which otherwise preserve $B-L$. However, we also point out that there are
many obstacles with this mechanism. Finally, we discuss baryogenesis by
assuming a singlet carrying a {\it global charge} and decaying mainly into SM
quarks and leptons to provide an adequate baryon to entropy ratio just at 
the end of reheating. Towards the end we conclude our paper by 
summarizing the facts.

\vspace*{-0.4truecm}
\section{Reheat temperature of the Universe}
\vspace*{-0.3truecm}
In models with a large extra dimensions, the reheat temperature 
is constrained 
from the possible thermal overproduction of gravitons in processes,
such as; $\gamma+\gamma\rightarrow G$, which requires 
$T_{\rm r}\ls 60$ MeV ~\cite{davidson0}. The second important observation
is that the inflaton field in these models has a natural coupling to the 
SM fields which is Planck mass suppressed \cite{abdel0}. This is  
due to the fact that the inflaton field resides 
in the bulk. This helps to inflate the size of the extra dimensions from 
its natural size; $(\rm TeV)^{-1}$ to its present millimeter size
in order to maintain the hierarchy, it also solves naturally 
the stabilization of the size of the extra dimensions \cite{abdel1}, 
and, besides all, it can provide an adequate density perturbations required 
for the structure formation in the Universe. As a consequence,
the inflaton has a  decay rate into 
Higgses, for instance, given as~\cite{abdel0}
 \be
 \label{decay} 
 \Gamma_{\phi\rightarrow {\rm HH}} \sim {g^2 M^3\over 32 \pi M_P^2} ,
 \ee
where $g$ the coupling constant,
$M$ is the fundamental scale which is related to the size of the extra
space; $V_n$, and, to the four dimensional Planck mass through~\cite{nima}
\begin{eqnarray}
M^{2+n}V_n=M_{\rm p}^2\,.
\end{eqnarray}
For $n=2$ extra dimensions $M$ can be at a TeV range. Current experimental
limits from collider physics and supernova 1987A imposes a bound;
$M\gs 30$ TeV~\cite{nima,exp}.

While deriving the decay rate in 
Eq.~(\ref{decay}), we have implicitly assumed that the mass of the inflaton
is roughly of the order of the fundamental scale $\sim M$,
in order to generate an adequate density fluctuations~\cite{abdel0,abdel1}. 
The estimated reheat temperature of the  Universe is given by 
$T_{\rm r} \sim 0.1\sqrt{\Gamma M_{\rm p}} \sim 1(10)$ MeV,
just right above the temperature required for successful Big Bang 
nucleosynthesis. It is also worth mentioning that the decay rate of 
the inflaton field into 
the relativistic particles, such as light degrees of freedom has a similar
suppression as Eq.~(\ref{decay}).  This is completely a different scenario
than that of the standard case where Planck scale is the fundamental scale.
For our case, the inflaton decay into the (non relativistic) 
Higgses is as favorable as
decaying into very light particles. This makes a difference while 
discussing the maximum temperature reached during the reheating era,
which is quite different from the reheat temperature of the Universe. 
As the inflaton field oscillates
with a decaying amplitude, the Universe is gradually filled up by the
light degrees of freedom which produces an effective temperature of the
Universe which follows a different scaling relationship between the 
temperature and the scale factor. The temperature reaches its maximum
when $a/a_{\rm I} \sim 1.48$, where $a$ denotes the scale factor 
of the Universe and the subscript ${\rm I}$ denotes the era when inflation comes
to an end. In the large extra dimension models, the inflationary scale
is determined by $H_{\rm I} \sim M$~\cite{abdel0,abdel1}. After reaching
the maximum temperature, it decreases as
$T \sim 1.3(g_{\ast}(T_{\rm m})/g_{\ast}(T))^{1/4}T_{\rm m}a^{-3/8}$,
where $T_{\rm m}$ denotes the maximum temperature~\cite{giudice,davidson1}. 
For $M\sim 10$ TeV,
the maximum temperature could reach $T_{\rm m} \sim 10^{5}$ GeV as mentioned
in Ref.~\cite{davidson1}. 
The basic assumption that goes behind this derivation
is that the inflaton field is predominantly decaying into
the relativistic species. However, this may not be  the case.
By reversing the argument, and, naively 
assuming that the inflaton decay populates only the non-relativistic 
degrees of freedom, one can show that the maximum temperature follows:
$M \gs T_{\rm m}\gg T_{\rm r}$, but, still much higher
than the reheat temperature of the Universe. 
Note in this case the  temperature-scale factor dependence, however, 
follows: $T\propto a^{-1}$.
Whatsoever be the case eventually the massive particles have to decay 
into a radiation bath, the decay rate of these intermediate particles are
now governed by their gauge couplings. If this happens the  
Universe might again be populated by radiation domination while the inflaton
field is oscillating. This could again raise the maximum temperature 
above $100$ GeV. Thus, the result apparently seems to be a robust one.
This might be a cheerful news for the electroweak baryogenesis.
However, it is still not clear whether the 
sphaleron transitions can be made useful for other sources of baryogenesis, 
such as leptogenesis. This is the topic we shall briefly meander upon  
before discussing the Affleck-Dine baryogenesis.
 
\vspace*{-0.4truecm}
\subsection{Leptogenesis}
\vspace*{-0.3truecm}
Following our previous discussion one might suspect that the lepton number
being produced in the decay process of a heavy fermionic singlet which carries
the lepton number, being processed into baryon number by anomalous  $B+L$
violating sphaleron interactions which are in equilibrium  for a temperature
more than $100$ GeV in the present circumstances. However, there is a simple
catch in this proposal. A singlet right handed neutrino can naturally couple to
the SM lepton doublet,  and, the Higgs field in a following way: $h \bar{L} H
N$~.  This leads to a potentially large Dirac mass term unless the Yukawa
coupling $h\sim 10^{-12}$, or, so. Moreover,  now the see-saw mechanism fails
to work, since, the largest  Majorana mass we may expect can never be larger
than  the fundamental scale. Therefore, given a neutrino mass  $\sim h^2
\langle H\rangle^2/M \sim h^2\cdot {\cal O}(1)$ GeV, we  still have to fine
tune $h^2 \ls 10^{-10}$, in order to obtain the right  order of magnitude for
the neutrino mass. Thus, the right handed neutrinos, if they at all exist,
are more likely to be  bulk fields rather than  brane fields. Since, in such
a case  the volume suppression of the bulk-brane coupling naturally provides  a
small coupling~\cite{nima2}. In any case  the  decay rate of the right handed
neutrino to the SM fields is suppressed by the smallness of $h$ that  gives
rise to a decay rate which is  similar to Eq.~(\ref{decay}). This 
 makes  extremely difficult to realize
baryogenesis, because, eventually when the right handed neutrino decays into
the SM fields, the background temperature is of the order  of the reheat
temperature, and, by this temperature the sphaleron  transition is not at all
in equilibrium. The sphaleron transition  rate is exponentially suppressed. So,
a seemingly suitable lepton number might not even get converted to the baryons
to produce the desired  baryon asymmetry in the Universe. 

On the other hand, it might be possible that sphalerons can reprocess a
pre-existing charge asymmetry into ba\-ry\-on asymmetry~\cite{olive} 
reflected in an excess of $e_L$ over anti-$e_R$ created during 
inflaton oscillations. This mechanism requires that $(B+L)$-violating 
processes are out of equilibrium before $e_R$ comes into chemical 
equilibrium, such that the created baryon asymmetry could be preserved. 
Again, this has to happen during, or, above $100$ GeV.  
Nevertheless, it is important to notice that still decaying 
inflaton field certainly injects more entropy to the thermal bath, provided
the inflaton dominantly decays into the relativistic degrees of freedom. 
So, an initially large baryon asymmetry has to be created in order
to obtain the right amount of asymmetry just before nucleosynthesis. 
One can easily estimate the amount of dilution that the last stages 
of reheating era will produce. The entropy dilution factor is given by :
 \be\gamma^{-1}=\left({s(T_{\rm r})\over s(T_{\rm c})}\right)
 =\left({g_{*}(T_{\rm r})\over g_{*}(T_{\rm c})}\right)
 \left({T_{\rm r}\over T_{\rm c}}\right)^3
 \left({a(T_{\rm r})\over a(T_{\rm c})}\right)^3~,
 \ee
where $s$ is the entropy and $T_{\rm c}$ denotes the electroweak temperature
$\sim 100$ GeV. For a low reheat temperature as $T_{\rm r} \sim 1$ MeV, the 
above expression gives rise to $\gamma^{-1} \gs 10^{25}$.
While calculating
the ratio between the scale factors, we have used $T\propto a^{-3/8}$.
Notice, that the lower bound appears, because, $g_{*}(T_{\rm c})
>g_{*}(T_{\rm r})$. This is due to the contribution coming from the heavier 
Kaluza-Klein (KK) graviton modes, albeit, their 
masses are much smaller than $T_{\rm c}$ that may be produced 
in the thermal processes, such as photon-photon fusion. Usually, these
heavy modes will decouple from the thermal bath right before nucleosynthesis. 
However, their contribution must be taken into account in the 
total relativistic degrees of freedom: 
$g_{*}(T_{\rm c}) = g_{*}(T_{\rm r}) + g_{*KK}$, which 
can be as large as the number of modes with masses between $T_{\rm c}$ and 
$T_{\rm r}$; thus giving, $g_{*KK}\ls R \Delta T\sim 10^{14}$, for
$R \sim {\rm mm}$. Strictly speaking the bound obtained on $\gamma^{-1} $
in our case is true only if $g_{*KK}=0$.
Therefore, including the entropy dilution factor, one concludes that the
initial  $n_{b}/s$ has to be extremely large $\gs 10^{15}$, in order to produce 
the required baryon asymmetry during nucleosynthesis, which is  
$n_{b}/s \sim 10^{-10}$. 
Such a large baryon  asymmetry is an  extraordinary requirement on any natural
model of baryogenesis, which is almost impossible to achieve in our case. 

There are couple of important lessons to be learned from the above analysis.
First of all the large production of entropy during the last stages of reheating
can in principle wash away any baryon asymmetry produced before electroweak
scale. The second point is that it is extremely unlikely that leptogenesis
will also work because one needs to inject enough lepton asymmetry in the
Universe before the sphaleron transitions are in equilibrium. The only
simple choice  left is to produce directly baryon asymmetry, however,
just before the end of reheating. The sole mechanism which seems to be 
doing well under these circumstances is the Affleck-Dine baryogenesis, 
which we shall discuss in the following section. 

\vspace*{-0.4truecm}
\section{Affleck-Dine Baryogenesis}
\vspace*{-0.3truecm}
Affleck and Dine have proposed a beautiful scenario of baryogenesis
in the context of supersymmetry \cite{affdine}. A scalar condensate which 
carries non-zero baryonic, or/and leptonic charge survives during inflation
and decays into SM fermions to provide a net baryon asymmetry. 
In our case the AD field; $\chi$, is a singlet carrying some global 
charge which is required to be broken dynamically in order to provide 
a small asymmetry in the current density. This asymmetry can be then
transformed into a baryonic asymmetry by a baryon violating interactions 
which we discuss later on. In order to break this $U(1)_\chi$ charge 
we require a source term which naturally violates $CP$ for a charged
$\chi$ field, and during the non-trivial helical evolution of the 
$\chi$ field generates a net asymmetry in $\chi$ over $\bar\chi$.
This necessarily has to happen after the end of inflation. 
Notice, that in our case the initial $CP$ phase is completely 
arbitrary and determined during the end of inflationary era.

We remind the readers that the inflaton energy density must govern the 
evolution of the Universe, and, the decay products of the inflaton is 
also responsible for reheating the Universe. This happens once
the inflaton decays before $\chi$ decays into SM quarks and leptos.
This decay of $\chi$ via baryon violating interaction generates 
a baryon asymmetry in the Universe which is given by 
 \begin{eqnarray}
 \label{ratio}
 \frac{n_{b}}{s} \approx \frac{n_{b}}{n_{\chi}}\frac{T_{\rm r}}
 {m_{\chi}}\frac{\rho_{\chi}}{\rho_{\rm I}}\,.
 \end{eqnarray}
The final entropy released by the inflaton decay is given
by $s \approx \rho_{\rm I}/T_{\rm r}$.  The ratio
$n_{b}/n_{\chi}$ depends on the total phase accumulated
by the AD field during its helical motion in the background of an oscillating 
inflaton field, which can at most be $\approx {\cal O}(1)$.
If we assume that the AD field is a brane-field,
then, the energy density stored in it can at most be: 
$\rho_{\chi} \approx m^2_{\chi}M^2$, on the other hand the energy
density stored in the (bulk) inflaton field is quite large $\rho_{\rm I}\
\approx M^2M_{\rm p}^2$~\cite{abdel0,abdel1}.
Thus, the ratio: 
$n_{b}/s \sim (T_{\rm r}/M_{\rm p})(m_{\chi}/M_{\rm p})\approx
10^{-34}(m_{\chi}/M)\ll 10^{-10}$, 
for $T_{\rm r} \sim {\cal O}(1-10)$ MeV. 
The conclusion of the above analysis is again disappointing, as it suggests
that the AD baryogenesis also leads to a small $n_b/s$. One way to boost this 
ratio is to assume that the AD field resides in the bulk. In that
case one naturally enhances the ratio $\rho_{\chi}/\rho_{\rm I}$,
however, keeping in mind that it is still less than one, in order not to 
spoil the successes of inflation.

Once, the AD field is promoted to the bulk, the energy density stored in the 
AD field rises to 
$\rho_{\chi} \sim m^2_{\chi}M_{\rm p}^2$~\cite{abdel0,abdel1},
this leads to the {\it maximum}  baryon to entropy ratio
\begin{eqnarray}
\label{final}
\frac{n_{b}}{s} \approx \left(\frac{T_{\rm r}}{M}\right)
\left(\frac{m_{\chi}}{M}\right)\, \sim 
10^{-10}\left(\frac{m_{\chi}}{{\rm 1 GeV}}\right) ,
\end{eqnarray}
where we have evaluated the right hand side for $T_{\rm r} \sim 10$ MeV and 
$M \sim 10$ TeV. Although, the mass of the AD field requires some 
fine tuning, up to the $CP$ phase, the above ratio can reach the 
observed baryon to entropy ratio quite  comfortably.  Notice, however, 
that  the actual predicted value  also depends on the initial conditions 
on $\chi$ that may render $m_{\chi}$ more freedom.  
Say for instance, if $\chi_0\sim M_{GUT}$, we get the right 
$n_b/s$ provided $m_\chi\sim M$.

We have noticed earlier that due to the violation of $U(1)_\chi$ charge, 
the dynamics of the  AD field generates an excess of $\chi$ over 
$\bar \chi$ fields. This asymmetry is transfered into baryon asymmetry 
by a baryon violating interaction, such as $\kappa\chi Q Q Q L$, 
however, keeping $B-L$ conserved. We also assume that $\chi$ interactions to
SM fields conserve $U(1)_\chi$ symmetry, thus, the 
quarks and leptons must carry a non zero global $\chi$ charge 
while the Higgs field does not. This avoids $\chi$
decaying into Higgses, which otherwise will reduce the baryonic abundance and
make the above interaction the main channel for its decay. 
While discussing the decay rate of $\chi$ field one has to  
take into account all 
possible decay channels which can be of the order of thousands  due to 
family and color freedom. On the other hand, 
we assume that the inflaton is decaying mainly 
into Higgses. 
Final result is then given by
 \be 
 \Gamma_\chi \approx  \left({\kappa\over g}\right)^2 
 \left({m_\chi\over M}\right)^7~\Gamma_{\phi} ~.
 \ee
By taking $\kappa/g\sim {\cal O}(1)$  we can insure
that $\chi$ will decay along with the inflaton, provided that its mass is 
very close to the fundamental scale. This will certainly demand some level of
fine tuning in the parameters.
We would like to mention that 
this is perhaps the simplest scenario one can think of for generating 
baryon asymmetry right before nucleosynthesis takes place.
It is worth mentioning that in our model 
the AD field will not mediate proton decay by dimension six operators as
$QQQL$, as long as $\chi$ does not develop any vacuum expectation value. 
Notice, other processes mediating proton decay, such  
as instanton effects might still occur.  While there is no known
solution for such a potential problem yet, our mechanism is at least not 
adding any new source to proton decay.
In the same spirit one may check those operators which induce $n-\bar n$
oscillations. Again, effective $\Delta B=2$ 
operators of dimensions 9; $UDDUDD$, and 11; $(QQQH)^2$,
can not be induced by integrating out $\chi$.

\vspace*{-0.4truecm}
\section{Conclusion}
\vspace*{-0.3truecm}

We have noticed that the observed baryon asymmetry in the Universe is difficult
to obtain in presence of a large extra dimensions. We have  pointed out that
there is a seemingly simple way, if we assume that there exists a  SM singlet
field carrying some global $U(1)_\chi$ charge which lives in the
bulk.  The non trivial dynamics of this field generates an asymmetry in
$\chi$-$\bar\chi$ after the end of inflation, which will be transfered 
into a baryon asymmetry by a baryon violating interaction. It is possible 
to insure that the AD field decays along with the inflaton such 
that the synthesis of the light elements can take place.

\vspace*{-0.4truecm}
\acknowledgements
\vspace*{-0.5truecm}
The authors are grateful to R. Allahverdi,  P. Binetruy, S. Davidson  and K.
Enqvist  for helpful suggestions.  A. M. acknowledges the support of  {\bf The
Early Universe network} HPRN-CT-2000-00152.

\vspace*{-0.5truecm}

\end{document}